\newcommand\PF[4]{{{}^{#1}\Psi^{#2}_{#3,#4}}}
\newcommand\cA{{\cal A}}
\newcommand\cB{{\cal B}}
\newcommand\cC{{\cal C}}
\newcommand\cH{{\cal H}}
\newcommand\cI{{\cal I}}
\newcommand\cV{{\cal V}}

\newcommand\AXB{{{}^{\cal A}\!X^{\cal B}}}

\documentclass[12pt]{article}
\usepackage{epsf}
\begin{document}

\begin{titlepage}
\nopagebreak
\begin{flushright}
June 2001\hfill
UT-943\\
hep-th/0106027
\end{flushright}

\renewcommand{\thefootnote}{\fnsymbol{footnote}}
\vfill
\begin{center}
{\Large Identity Projector and D-brane}\\
{\Large in String Field Theory}

\vskip 20mm

Yutaka MATSUO\footnote{
{\tt E-mail address: matsuo@phys.s.u-tokyo.ac.jp}
}
\vskip 1cm
Department of Physics, University of Tokyo\\
Hongo 7-3-1, Bunkyo-ku\\
Tokyo 113-0033\\
Japan
\end{center}
\vfill

\begin{abstract}
We study the identity projectors
of the string field theory in the generic BCFT background.
We explain how it can be identified as the projector
in the linking algebra 
of the noncommutative geometry.
We show that their (regularized) trace is exactly given by
the boundary entropy which is proportional to
the D-brane tension.
\end{abstract}

\vfill
\end{titlepage}

In the previous letter \cite{r-Mat2}, 
we argued that there are two types of
the projectors for each of the Cardy state
\cite{r-Cardy}
in the cubic open string field theory
\cite{r-Witten}
in the generic BCFT background. One is the sliver 
state discussed in many articles recently 
\cite{r-RZ, r-KP, r-RSZ, r-RSZ1, r-RSZ2, r-GT, r-RSZ3}
in the context of the vacuum string field theory.
It is conjectured to describe the D-brane in the sense
that it reproduces the D-brane tension 
in the level truncation \cite{r-RSZ1} or 
the renormalization group flow \cite{r-RSZ3} methods.

In this letter, however, we study the
other projector which is the generalization of the
identity operator.  We are interested in this operator
since it is interpreted as
the nontrivial projector in the {\em linking algebra} \cite{r-RW, r-KMJ}
which appear in the general framework of the noncommutative 
geometry \cite{r-CDS, r-SW}. 
In short, the linking algebra is a huge algebra of the 
open string fields with the star multiplication
which link any pair of all D-branes which is possible from
the bulk CFT.  
The detail will be explained in the text.

The identity operator also has a benefit 
that some exact calculations are possible for the
topological invariants 
in the sense of the operator algebra K-theory
\cite{r-Connes, r-WO, r-Matsuo, r-HM}.
In particular the first nontrivial example, the trace of
the projector, is exactly given as the
boundary entropy \cite{r-AL} (see eq.(\ref{e-trace2})).
This quantity was identified with the D-brane tension in
\cite{r-HKMS}. It also appeared in the context of
the boundary string field theory \cite{r-KMM,r-deAlwis}.
It gives an explicit and analytic proof 
of the conjecture in \cite{r-RSZ1} that the trace of the projector
is proportional to the brane tension.  In this sense,
the identity projector is at least an equally legitimate candidate
which represents the D-brane as the sliver projector.

As explained in the previous letter \cite{r-Mat2},
our goal is to extend Witten's conjecture
\cite{r-Witten1,r-Witten2} that the D-brane charge is classified by
the K-theory to the operator theory of the conformal field theory.
We hope that the result in this letter is a relevant step
toward that direction.

We start from reviewing some of the essential features of the
noncommutative soliton \cite{r-GMS} in rather abstract
language for the application to the string field theory algebra.
In the noncommutative geometry, the commutative ring of the
continuous functions over some topological space $X$ is replaced
by the noncommutative $C^*$-algebra $\cA$.  The simplest example
is the Moyal plane where $\cA$ is given by the set of the 
bounded linear operators acting on the Hilbert space of the
harmonic oscillators $\cal B(H)$. The noncommutative soliton is defined as the
projector in this algebra,
\begin{equation}\label{e-proj}
 p\in \cA,\quad p^2=p,\quad p^*=p.
\end{equation}
For the Moyal plane, the rank of the projector is
interpreted as the soliton number,
\begin{equation}
 \tau(p)=\mbox{Tr}_\cH \  p=n\in {\bf Z_{\geq 0}}
\end{equation}
For the application to the D-brane \cite{r-HKLM},
$\tau(p)$ is identified with the number of D-branes
created after the tachyon condensation \cite{r-Sen}.
However, for the more general noncommutative space, the trace
is not always quantized and may take a continuous spectrum.
This is, actually, the major difference between the topological
and the operator theoretical K-group.
For example, in the noncommutative torus with the irrational
noncommutative parameter $\theta$, the spectrum becomes 
dense between 0 and 1 with the following form \cite{r-BKMJ,r-KMJ},
\begin{equation}\label{e-torus}
 \tau(p)=n+m\theta,\quad
 n,m\in {\bf Z},
\end{equation}
and it represents the composite system of $n$ D2 and $m$ D0
branes.

In the application to the string theory, this type of the projector
is used to describe the lower dimensional D-brane charges
which appear after the tachyon condensation \cite{r-Sen}
of the original D-brane described by the algebra $\cA$.
In the following, we argue that there are another type of the
projectors which appears naturally in the open string field theory.

In general the open string may attach its
two ends on the different D-branes which
have the different type of the noncommutative geometries.
To describe such a situation, we need to consider the
two D-branes (or the two $C^*$-algebras) as a whole.
For that purpose, it is essential to
introduce the notion of the Morita equivalence
bimodule \cite{r-CDS,r-SW,r-Schwarz}. 
Two noncommutative geometries 
described by algebras $\cA, \cB$ is called Morita equivalent
if there exists a bimodule $\AXB$ where $\cA$ (resp. $\cB$)
acts from the left (right) on $\AXB$.  
In the string theory, this bimodule is identified
with the open string that interpolates the different 
D-branes.  In this bimodule we need to have
$\cA$- ($\cB$-) valued inner product $(,)_\cA$ ($(,)_\cB$)
on $\AXB$. They need to satisfy,
\begin{eqnarray}
 (ax,y)_\cA&=&a(x,y)_\cA, \quad  (x, \bar{a} y)_\cA=(x,y)_\cA a\nonumber\\
 (x,yb)_\cB&=&(x,y)_\cB b, \quad  (x\bar{b}, y)_\cB=b(x,y)_\cB\nonumber\\
 (x,y)_{\cA}z&= & x (y,z)_\cB
\label{e-associativity}
\end{eqnarray}
where $x,y,z\in \AXB$ and $a\in \cA$, $b\in \cB$.
The string interpretation of these inner products is
the star product of the string fields between
the interpolating strings. We illustrate it in the Figure 1.
\begin{figure}[th]
\centerline{\epsfxsize=12cm \epsfbox{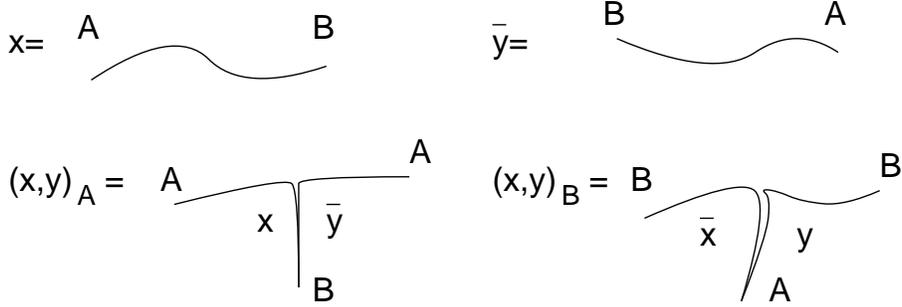}} \vskip 3mm
\caption{String interpretation of inner product of the bimodule}
\end{figure}
The equations (\ref{e-associativity}) 
are naturally interpreted as  the associativity of the star product.
We note that these abstract relations are also practically
important to construct the noncommutative soliton \cite{r-MM, r-KMJ}.

With the help of the Morita bimodule, one may define a
new $C^*$-algebra $\cC$ (which is called the {\em linking algebra}
\cite{r-RW}),
\begin{equation}
 \cC=\left\{\left.\left(
\begin{array}{c c}
 a & x\\ \bar{y} & b
\end{array}
\right)\right| a\in \cA,\,\, b\in \cB,\,\, x,y\in \AXB \right\}\,\,.
\end{equation}
The bar in $\bar{y}$ is the involution which turns the 
element of $\cA$-$\cB$ bimodule into the $\cB$-$\cA$ bimodule.
This algebra can be interpreted as the
open string system in the presence of two D-branes specified by
$\cA, \cB$ \cite{r-KMJ}. 
In this composite system, we may define
two obvious projectors as,
\begin{equation}\label{e-proj2}
 p =\left( \begin{array}{c c}
     1_\cA& 0 \\ 0 & 0
	   \end{array}\right),\quad
 q =\left( \begin{array}{c c}
     0 & 0 \\ 0 & 1_\cB
	   \end{array}\right),\quad
\end{equation}
which satisfies $p\cC p=\cA$, $q\cC q=\cB$ and $p+q=1_\cC$.
In this sense, $p$ and $q$ may be interpreted as the projectors
to the algebras $\cA$ and $\cB$. They project into the single
D-brane algebras from the composite system.

As mentioned before, these projectors are 
slightly different from those appearing in
the noncommutative soliton (\ref{e-proj}).  
Firstly the projectors
in (\ref{e-proj2}) define the projections
from the composite system of two D-branes but not from a single D-brane.  
Secondly, while the rank of the projector in (\ref{e-proj})
is usually finite, that of (\ref{e-proj2}) may be
infinite since it represents the rank of the whole 
algebra $\cA$ (or $\cB$).

One may straightforwardly extend the notion of the linking
algebra  to the open string
system with $N$ D-branes.  In this case, the linking
algebra $\cC$ is described by $N$ by $N$ matrix
and we have $N$ projectors which generalizes $p$ and $q$.

Let us proceed to apply this abstract idea to the  
more concrete setup of the BCFT.
As we briefly reviewed in our previous letter\footnote{
We use the notation of \cite{r-BPPZ} for the BCFT in this letter.},
the general strategy of the construction of the D-brane
(the Cardy state) is as follows.
Starting from the bulk CFT that defines the closed string,
we have the algebraic data $\cI$ which represents the set
of the chiral primary fields $\phi_i(z)$ ($i\in\cI$).
For each of the chiral primary fields, there exists
a closed string state (the Ishibashi state \cite{r-Ishibashi})
$|i\rangle\!\rangle$. These states themselves, however, do not
define the consistent boundary in the open string sector.
For that purpose, we need to take their linear combination,
\begin{equation}\label{e-Cardy}
 |a\rangle = \sum_j \frac{\psi^j_a}{\sqrt{S_{1j}}}|j\rangle\!\rangle
\end{equation}
such that it reproduces the integer number of the chiral
primary field in the open string
channel as (after the modular transformation),
\begin{equation}\label{e-Cardy2}
 \langle b|\tilde{q}^{\frac{1}{2}(L_0+\bar L_0-\frac{c}{12})}|a\rangle
 =\sum_i \chi_i(q) n_{ia}{}^b\,\,,
\end{equation}
with $q=e^{2\pi i \tau}$  and $\tilde{q}=e^{-2\pi i/\tau}$.
The constraint is that the coefficients,
\begin{equation}\label{e-nia}
 n_{ia}{}^b=\sum_j \psi^j_a (\psi^j_b)^{*} \frac{S_{ij}}{S_{1j}}.
\end{equation}
should be non-negative integers.

We suppose that the set of the Cardy state is labeled by
$a\in \cV$.  In the rational conformal field theory, the set 
$\cV$ is the finite set but in general we have the infinite 
number of the consistent boundary states.
The boundary field $\PF{a}{b}{i}{\beta}(x)$ which represents 
the open string has four labels, $a,b\in \cV$ for the two boundaries,
$i\in \cI$ for the chiral primary field and finally $\beta=1,\cdots,
n_{ia}{}^b$ for each channel. We write the highest weight state
associated with this field as $|a,b\,;\,i\,;\,\beta\!\!>$.

In our previous letter \cite{r-Mat2}, we proposed
the string field theory algebra from these data.
It is actually the linking algebra $\cC$ defined as follows.
Suppose we start from the D-brane system of $N_a$ D-branes
of type $a$. The size of the linking algebra becomes $N\times N$
with $N=\sum_{a\in\cV} N_a$. We label it by a pair of indices
$((a,I_a),(b,I_b))$ with $I_a=1,\cdots,N_a$
or by $(A,B)$ $A,B=1,\cdots,N$ if we want to abbreviate it. 
The label $I_a$ are Chan-Paton index.
The operators in $(A,B)$th 
entry belongs to the set of the states of the form,
\begin{equation}
\left\{L_{-n_1}\cdots L_{-n_\ell} |A,B\,;\,i\,;\,\beta\!\!>\,
\left|\quad
i\in \cI, \,\beta=1,\cdots,n_{ia}{}^b
\right.\right\}\,\,.
\end{equation}

We denote the set of the Hilbert space as $\cH_{A,B}
=\cH_{(a,I_a)(b,I_b)}$.
The star product in the string field theory
is defined as the mapping,
\begin{eqnarray}
 \star & : & \cH_{AB}\otimes \cH_{BC} \rightarrow \cH_{AC}\nonumber\\
 && |v\!> \otimes |w\!> \rightarrow |v\star w\!>
= \mbox{bpz} (\langle V_3| |v\!>\otimes |w\!>)\,,
\end{eqnarray}
where $\langle V_3|$ is the Witten's 3 string vertex operator
and bpz is the BPZ conjugation. In \cite{r-Mat2} we argued that
this algebra is in principle computed by the knowledge of the
OPE of the boundary fields,
\begin{equation}\label{e-prod}
 {}^b\Psi^c_{i,\alpha_1}(x_1) {}^c\Psi^a_{j,\alpha_2}(x_2)
 =  \sum_{p,\beta, t} {}^{(1)}F_{cp}\left[
\begin{array}{c c}
 i&j\\ b & a
\end{array}\right]^{\beta\ t}_{\alpha_1\ \alpha_2}
\frac{1}{x_{12}^{\Delta_i+\Delta_j-\Delta_p}}
 {}^b\Psi^a_{p,\beta}(x_2)+\cdots,
\end{equation}
and the conformal Ward identities derived in \cite{r-RZ}
for the three string vertex. 
This linking algebra can be decomposed into the diagonal pieces
$\cH_{AA}$ which give the algebra on the D-brane $a$
and the off-diagonal pieces $\cH_{AB}$ which define
the Morita bimodule.  As long as the label $A$ has of the 
form $(a,I_a)$ with the same $a$, the diagonal algebra $\cH_{AA}$ 
are isomorphic with each other
and we write it as $\cA^a$.

One of the critical step here is to observe that
the $C^*$-algebra $\cA^a$ is unital for any Cardy state $|a\rangle$.
This fact was implicitly written in \cite{r-RZ} and explicitly
stated in \cite{r-Mat2}.
The identity operator $\cI^a$  ($a\in\cV$) was defined
through the identity chiral
field in each $a$-$a$ sector and the global conformal transformation
\cite{r-RZ}
\begin{equation}
 w=\left(\frac{1+iz}{1-iz}\right)^2.
\end{equation}
Explicitly an elegant expression was found in \cite{r-EFHM},
\begin{eqnarray}
 \cI^a &=& \left(\prod_{n=2}^\infty \mbox{exp}
\left\{-\frac{L_{-2^n}}{2^{n-1}}\right\}
\right)e^{L_{-2}}|a,a;0\!>\nonumber\\
 & = & \ldots e^{-\frac{1}{2^2}L_{-2^3}}\,
e^{-\frac{1}{2}L_{-2^2}}\,
e^{L_{-2}}|a,a;0\!>.
\end{eqnarray}
From this operator, one may define the 
the analogue of $p$ and $q$ in (\ref{e-proj2}).
We use the similar symbol $\cI^A$ as the projector
where the identity operator $\cI^a$ is located at $AA$-th entry
($A$ is of the form $(a,I_a)$) and zero elsewhere.
They satisfy
\begin{equation}
 \sum_{A=1}^N \cI^A=1_\cC,
\quad \cI^A\star \cI^B =\delta_{AB}\cI^B,\quad
 \cI^A \star \cC \star \cI^A \sim \cA^a\,\,,
\end{equation}
or
\begin{equation}
 \cI^A \star |B,C\!> =\delta_{AB} |B,C\!>,\quad
 |B,C\!>\star \cI^A  =\delta_{CA} |B,C\!>
\end{equation}
for any $|B,C\!>\in \cH_{BC}$ embedded into the $BC$th entry.  
As we discuss in the definition
of the linking algebra, we would like to interpret this projector
as characterizing each D-brane.

In the following, we would like to calculate the trace of 
$\cI^a$. 
Usually it has been argued that trace
diverges and such kind of the calculation
is meaningless. There is, however, a systematic
method to regularize the infinity and we conjecture 
that the finite outcome will have the meaningful interpretation.

Since $\cI^a$ is the identity element in the algebra $\cA^a$,
we first evaluate the number of the generators of $\cA^a$
which is identified with the number of states
$|\cH_{aa}|$.
(For the simplicity 
we omit Chan-Paton indices and replace the index $A,B,\cdots$
by Cardy state index $a,b,\cdots$.)
As in the calculation of the character, we regularize
the counting of the states by introducing $\tilde{q}^{L_0-c/24}$ 
with $\tilde q=e^{-2\pi i/\tau}$ and take the
$\tilde{q}\rightarrow 1$ ($\tau\rightarrow i\infty$) limit later.  
\begin{eqnarray}
 |\cH_{aa}|&=&\lim_{\tilde q\rightarrow 1} \mbox{Tr}_{\cH_{aa}} 
\tilde q^{L_0-c/24}
= \lim_{\tilde q\rightarrow 1} \sum_i n_{ia}{}^a \chi_i(\tilde q)\nonumber\\
& = & \lim_{q\rightarrow 0}
\langle a| q^{\frac{1}{2}(L_0+\bar L_0)} |a\rangle\nonumber\\
&= & |\langle \mbox{vac} |a\rangle|^2\,\,,
\end{eqnarray}
with $q=e^{2\pi i \tau}$.
Here $\chi_i(q)$ is the character for the chiral primary field $\phi_i$
and $\langle\mbox{vac}|$ is the closed string vacuum state.
In passing from the first to the second line, we used 
eq.(\ref{e-Cardy2}).  The final answer becomes finite!
Indeed this type of the calculation is typical in the
BCFT, for example, in the norm calculation of the Ishibashi state 
(see for example \cite{r-BPPZ}).  

We have to be careful not to confuse this quantity as the trace of
$\cI^a$.  In the finite $N\times N$ matrix case, the number of the
generators of the algebra is $N^2$ but the trace of the identity
operator is $N$.  While it may be naive to apply it to 
the infinite $N$ case, we 
have to use this analogy to conclude the trace
formula for $\cI^a$,
\begin{equation}\label{e-trace2}
 \tau(\cI^a) = \langle \mbox{vac}|a\rangle.
\end{equation}
As we mentioned at the beginning of this letter, this
quantity is called the boundary entropy \cite{r-AL}
and gives the tension of the D-brane \cite{r-HKMS,r-KMM, r-deAlwis}.  
We think that
this result  is quite satisfactory and one may safely conjecture that
the identity projector is a candidate
to represent the D-brane in the open string field theory.

More generally the projection $p$ to the composite system
of $n_a$ D-branes of the type $a$ has the trace,
\begin{equation}
 \tau(p)=\sum_{a\in\cV} n_a \langle \mbox{vac} | a\rangle
\end{equation}
which may be regarded as the generalization of the formula
for the quantum torus (\ref{e-torus}). The integrability of the
D-branes appear in the coefficients but $\tau(p)$ as a whole
is not necessarily the multiple of the integer.


To conclude this letter, we would like to give some comments.
\begin{enumerate}
 \item We summarize the reason why we believe that
       the rank infinite projector is more natural than
       the rank finite ones.
       Suppose we have a projector $p$ which describes a
       D-brane system. It gives a subalgebra of the original open string
       field theory algebra $\cC$ in the form $\cA=p\,\cC\, p$.
       If $p$ is the rank finite projector, $\cA$ is isomorphic
       to the finite size matrix algebra and should be identified
       as the group of the Chan-Paton factor.   On the other 
       hand, if we use our infinite rank projector, $\cA$ is the subalgebra
       of $\cC$ generated by
       infinite number of the open strings whose two ends are
       attached to the given D-brane we would like to pick up.
       We think that the rank finite algebra is too simple to be
       identified as the algebra generated by the open string fields.
 \item We have pointed out that there are some differences
       between the conventional noncommutative soliton and
       the projectors in the linking algebra.  This difference,
       however, should be superficial.  
       If the D-brane in the higher dimension can be decomposed
       into the collection of the lower dimensional D-branes, it simply
       implies that the Cardy state associated with the higher dimensional
       brane is reducible and can be decomposed into the sum of the
       irreducible ones,
       \begin{equation}
	|a\rangle = |a_1\rangle + |a_2\rangle +\cdots.
       \end{equation}
       In such a situation, the $C^*$-algebra of the higher dimensional
       D-brane is actually considered as the linking algebra of the
       lower dimensional D-branes and the projectors of the both
       approaches become the identical.  It is of some interest to check
       that such a decomposition is possible in the tachyon condensation
       of the noncommutative D-branes \cite{r-HKLM}.
       A plausible answer is that the algebra of the open string fields
       of $D$ $(p+2)$-brane can be factorized as $\cC={\cal B(H)}
       \otimes \cA$ where $\cA$ is the algebra of $D$ $p$-brane.
       If this guess is true, the rank finite projector $p$
       in \cite{r-HKLM} should be mapped to
       the rank infinite projector
       of the string fields as $p\otimes 1_\cA$.
       
 \item From the viewpoint of the noncommutative geometry, 
       the quantity (\ref{e-trace2}) is the simplest 
       geometrical invariant made from the
       projector.  The general invariants are constructed by 
       pairing with the
       cyclic cohomology element $\tau_n$ \cite{r-Connes}
       \begin{equation}
	\tau_n(\cI^{a},\cdots, \cI^{a}).
       \end{equation}
       We would like to know how the notion of the cyclic cocycle
       can be extended to the string field theory algebra.
       In the case of the commutative situation, such invariants
       are expressed by using the differentiation of the projector.
       In the operator language, we need to use the operator
       product expansion (\ref{e-prod}) to express such operation.
       It seems to us an  essential step to seek such a possibility
       to explore the K-group of the string field theory algebra
       and the D-brane charge \cite{r-Witten1}.
\end{enumerate}

\vskip 10mm

\noindent{\sl Acknowledgement:}
The author would like thank T. Eguchi, T. Kawano, 
T. Takayanagi and K. Ohmori for the useful conversations.

The author is supported in part by Grant-in-Aid (\#13640267)
and in part by Grant-in-Aid for Scientific Research
in a Priority Area ``Supersymmetry and Unified Theory of 
Elementary Particle'' (\#707) from the Ministry of Education,
Science, Sports and Culture.


\end{document}